# Adapting Blockchain Technology for Scientific Computing

Wei Li
weili@weililab.org

**Abstract**
Blockchain stores information into a chain of "blocks", whose integrity is usually guaranteed by Proof of Work (PoW). In many blockchain applications (including cryptocurrencies), users compete with each other to win the ownership of the blocks, a process commonly referred as "mining". Mining activities consume huge amount of power, while the outcome appears to be useless besides validating a block. Here we discuss the requirements of designing a new PoW algorithm. We also propose a PoW scheme to help solve high-dimension, non-linear optimization problems. Simulation experiments of blockchains generated by three miners solved an instance of Traveling Salesman Problem (TSP), a well-known NP-hard problem. The revised scheme enables us to address difficult scientific questions as a byproduct of mining.

**Introduction**
Blockchain is a de-centralized, peer-to-peer network of a ledger system that is inherently resistant to modification. Blockchain stores information (e.g., transactions) into a data structure called "blocks", that are inter-connected as chains or networks [1]. To prove the integrity of a block and to increase the difficulty of modification and attack, each block usually includes proof-of-work (PoW), whose generation requires some computational resources. PoW is widely used in cryptocurrencies including BitCoin [1] and Ethereum [2]. The increasing popularity of cryptocurrency stimulates users to use dedicated hardware and consume huge amount of electricity to compete for the ownership of the block ("mining"). However, the outcome, which is a hash value, is considered as useless beyond PoW.

We set out to study the following question: can PoW be used for scientific computation to answer interesting and important questions? Among different types of cryptocurrencies, Gridcoin is proposed to reward uses who contribute to Berkeley Open Infrastructure Network Computing (BOINC) Grid [3], and CureCoin rewards users to perform protein folding calculations by joining Folding@home network [4]. However, both approaches distribute coins to users from the statistics of some special network and is not generalizable to other blockchain applications. Other primitive approaches are proposed to process NP-complete problems or imaging processing algorithms ([5], [6]).

Here we discuss the general requirements of a new PoW algorithm. Based on these requirements, an alternative PoW is proposed to incorporate scientific computing. The revised PoW asks users to optimize a high-dimensional, non-linear objective function, while the difficulties are adjustable. Therefore, multiple rounds of competition of PoW in blockchain applications help us solve scientific problems that generally require super-computers. We present the simulation results of three miners whose PoW is used to solve a NP-hard Traveling Salesman Problem (TSP).

**Introduction of PoW (Proof-of-Work)**
Each block in a blockchain includes three components: validation information from previous block, information to be stored in the current block, and PoW. For example, each block in Bitcoin contains the hash value of the previous block (*prev_block_hash*), records of transactions, and a specific integer value (*nonce*). Transactions are stored in a Merkle tree structure whose hash value is calculated (*transaction_hash*). User need to find the value of *nonce* such that the hash output from the three components is smaller than certain value. The mining algorithm of Bitcoin can be summarized as follows:

> ***Mining(prev_block_hash, transaction_hash, target):***
>   ***For*** *nonce in 0 to Inf:*
>     ***If*** *hash_output(prev_block_hash, transaction_hash, nonce)< target:*
>       *Announce the success of mining*

The value *target,* shared across all miners, determines the difficulties of mining and are adjusted from time to time. Users compete with each other to win the ownership of the block. The structure and the PoW process can be illustrated in Figure 1.

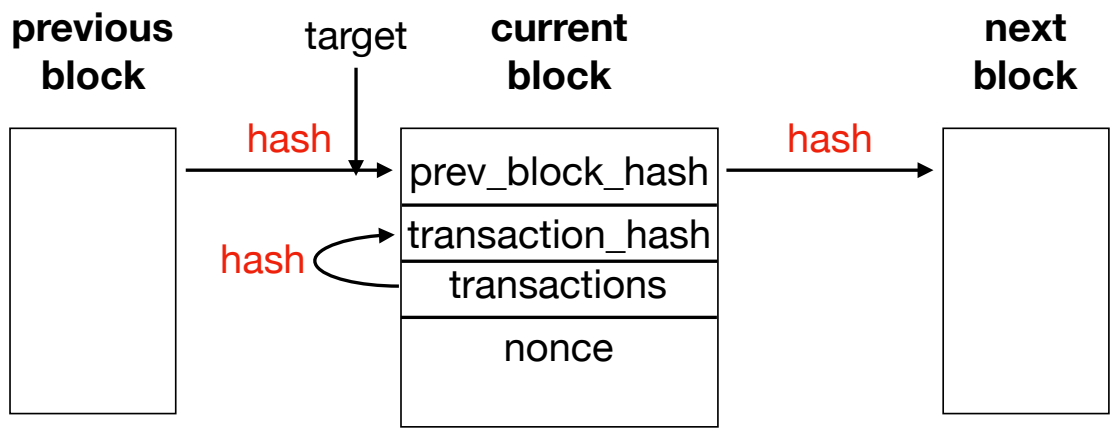

**Figure 1**. A block and the PoW process in bockchain.

The hash function has the following properties to be used for PoW:

1. Difficult to find a value to meet certain condition (i.e., smaller than a threshold);
2. Easy to validate;
3. Block information (like transaction) can be integrated;
4. The difficulty can be adjusted.

Conditions 1 is needed to increase the difficulty of future modification and attacks from hackers. 2 is also needed since all users of blockchain need to validate the proposed PoW. 3 is implemented to avoid pre-computing of PoW: users compute results before block generation to gain advantages over others. 4 is required to adjust for different number of competitors.

In the hash-based PoW, 1 and 2 is satisfied since finding a nonce whose hash value is small enough is difficult, but validating it is easy. The information of the transactions, *transaction_hash,* is considered during the PoW (condition 3), and the value of *target* changes from time to time, reflecting the difficulty of mining (condition 4).

We set out to find another function to replace hash function and to satisfy conditions 1-4.

## A PoW Algorithm for Scientific Computing

Many scientific and engineering problems set out to find a minimum (or maximum) value of a certain objective function (e.g., energy, entropy). Generally, we need to find a value $\theta$ in a N-dimension parameter space $\Theta = R^N$ such that the objective function $f$ is minimized:

$$\min f(\theta), \theta \in \Theta$$

The value of N may be large and $f$ may be non-linear, non-concave. Therefore, finding (and validating) the *global minimum, or even the local minimum* of $f$ is difficult. However, given a value $\theta$ and $c$, validating whether $f(\theta) = c$ is easy. Validating whether $c$ is a *local minimum* is relatively easy: we could randomly sample $\theta'$ who is close to $\theta$, and check whether $f(\theta') < c$. The more samples $\theta'$ we check with $f(\theta') > c$, the more likely that $c$ is a *local minimum*. Therefore, it is possible to use $f$ for PoW, as conditions 1 and 2 are met. For blockchain applications, the validation of *local minimum* can further be performed simultaneously in different users, each performs independent validation from randomly generated values of $\theta'$.

We now design an algorithm to satisfy conditions 3 and 4. The algorithm works as follows: for large N, only perform a *local minimum* search in K coordinates of $\theta$, while the rest N-K coordinates are unchanged. The value of K, or the degree of freedom, can be adjusted for difficulty. For K=1, the optimization process is easy (as search is performed in only 1 coordinate). The difficulty of finding a local minimum increases exponentially as K increases, a phenomenon referred as "combinatorial explosion". It becomes the most difficult when K=N, where the *local minimum* may become *global minimum*. To incorporate block information, we design an index function to determine which K out of N dimensions are chosen for optimization, based on the hash values of the block.

Here is a formal definition. Let $\theta = (\theta_1, \theta_2, \ldots \theta_N) \in \Theta = R^N$. S is a set of dimension indices defining a K-dimension subspace out of $\Theta$:

$$S = \{S_1, \ldots, S_K\}, S_1, \ldots, S_K \in \{1, 2, \ldots, N\}$$

$\theta^* \in \Theta$ is a *bounded local minimum (BLM)*, if

$$f(\theta^*) < f(\theta),$$

for all $\theta \in \Theta$ satisfying the following conditions:

$$\theta_i = \theta_i^*, i \notin S$$
$$|\theta_i - \theta_i^*| < \delta, i \in S \qquad (1)$$

where $\delta$ is a pre-defined parameter to define the size of neighbor points near $\theta^*$ (e.g., $0.01 * |\theta_i^*|$). We write

$$\theta^* = BLM_f(S)$$

if $\theta^*$ is a bounded local minimum.

Therefore, for a difficulty threshold *K*, we can change the PoW algorithm as follows.

```
Function Mining(prev_block_hash, transaction_hash, K):
  S=index(hash_output(prev_block_hash, transaction_hash), K)
  For θ ∈ Θ:
    If a BLM is found: θ* = BLM_f(S):
      Announce the success of mining
  If others announce the success of mining (θ* = BLM_f(S')):
    Check if θ' that is close to θ* satisfy f(θ*) < f(θ')
```

A description of the PoW is shown in Figure 2. For a large value of N and a small value K, the number of possible K combinations may be huge. For example, For N=100 and K=10, the possible combinations are *combination(100,10)*=17,310,309,456,440. Therefore, it's unlikely that two users will perform optimization in a same K-dimension parameter subspace.

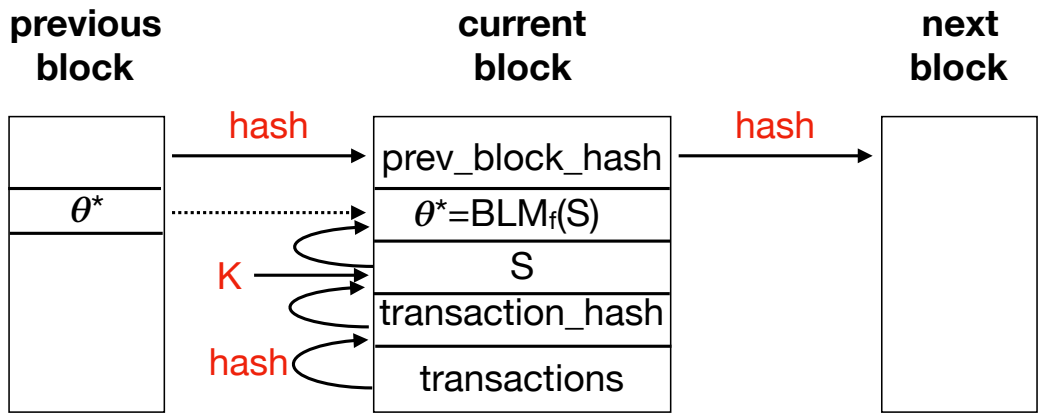

**Figure 2.** The improved PoW.

For each round of block generation, PoW can further starts from the $\theta^*$ value generated in previous block (dash arrows in Figure 2). This enables a continuously minimization of $f$ in a series of blocks. To avoid deadlock, restart the optimization process by generating a new random $\theta$.

**Index Function**
The index function takes the hash value (calculated from previous block and transaction records) as input and generates a K-dimension subspace from N dimensions. This can be done by iteratively performing mod operations on the hash value:

```
Function index(hash_value, K):
  S = {}
  While |S| < K:
    T = hash_value mod N
    If T ∉ S:
      S = S ∪ {T}
    hash_value=hash_value/N
```

In the iteration, if *hash_value* becomes zero before K values are extracted, we can use *hash_output(hash_value)* (and iteratively hashing, if necessary) to generate a new hash value, until K non-repetitive indices are chosen.

**Adjusting Mining Difficulty**
As is mentioned above, the value K adjusts the difficulty of mining activities: a higher K value makes it difficult to search for *local minimum*. We can also adjust the difficulty by setting up the number of *significant figures* of $\theta^*$, and the values of $\delta$ in Equation (1). For example, we require $\theta^*$ to have at least K significant figures and set $\delta = K/N * |\theta_i^*|$. For higher K values, users need more computational resources to get a more precise value of $\theta^*$ (e.g., $\theta^* = 3.2e3$ for K=2 and $\theta^* = 3.21543e3$ for K=6), and guarantee a larger window defined as *local minimum* (for N=100, $\delta = 0.02|\theta_i^*|$ for K=2 and $\delta = 0.06|\theta_i^*|$ for K=6).

Note that the Bounded Local Minimum (BLM) is different from the typical definition of *local minimum*, in that the latter requires $\delta \to 0$. A large value of $\delta$ in BLM prevents users from just searching for trivial *local minimum* in some cases. Consider a simplified example for K=1, where the objective function is $y = \sin(x) + 0.1\sin(20x), x \in [0, 2\pi]$. The values of x and y are plotted in Figure 3. Some local minimum (blue dots) will not be accepted as it is not BLM for a larger window. Some local minimum may still be accepted for smaller window (purple dots)

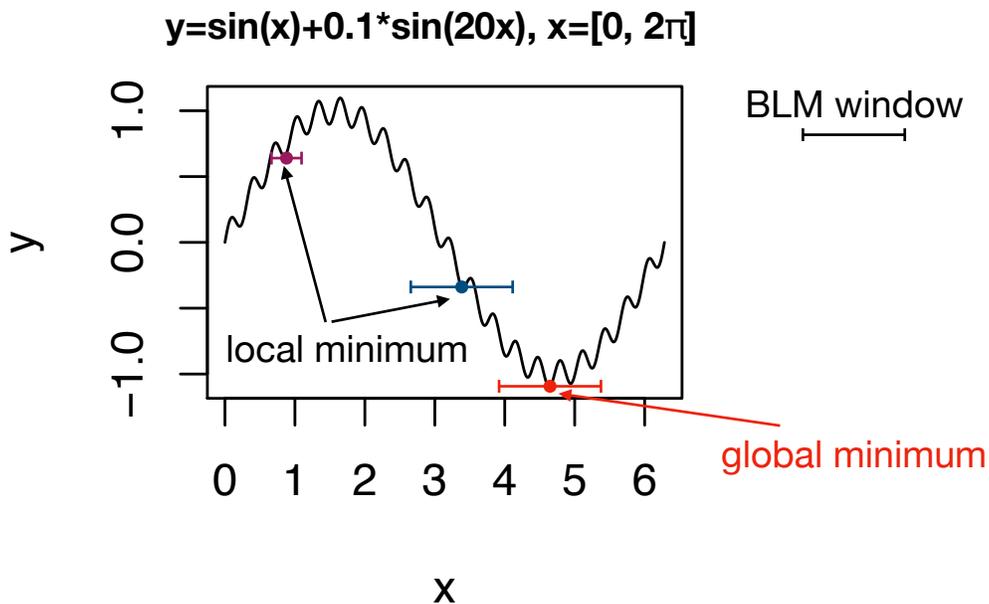

**Figure 3.** BLM is different from a typical local minimum.

**Simulated Blockchain Experiments for Traveling Salesman Problem (TSP)**
Traveling Salesman Problem (TSP) seeks the shortest route between cities that visits each city exactly once and returns to the original city. This is a well-known NP-hard problem, and a dynamic programming based Held-Karp algorithm finds the exact solution in $O(N^2 2^N)$ time for $N$ cities. An example of a TSP problem is provided in Figure 4. Besides the exact algorithm, various approximation algorithms exist.

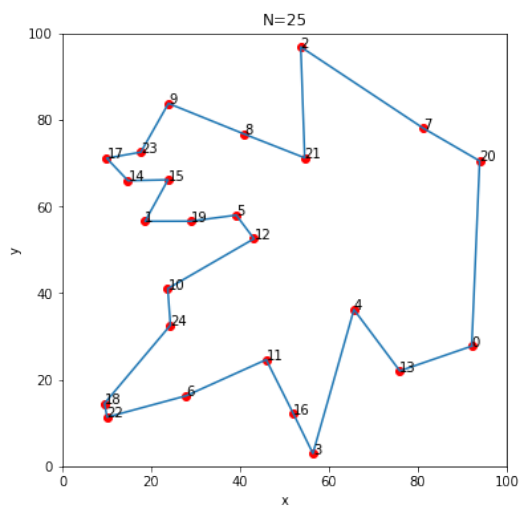

**Figure 4: A solution to the TSP problem with 25 cities (N=25). The 2-d coordinates of these cities and the exact solution using Held-Karp algorithm are shown.**

We set out to use the proposed PoW scheme to solve a TSP problem. First, we need to define what the K-dimension subspace (S) are, how to find the neighbor points in (1), and the way to adjust mining difficulty. For $N$ cities with index $0, 1, \ldots, N-1$, a loop route $\theta = (0, \theta_1, \theta_2, \ldots \theta_N, 0)$ starts from city 0, goes through cities $\theta_1, \theta_2, \ldots \theta_N \in \{1, \ldots, N-1\}$ and ends with city 0.

For a K dimension subspace $S = \{S_1, \ldots, S_K\}, S_1, \ldots, S_K \in \{1, 2, \ldots, N-1\}$, the corresponding loop route $\theta = (0, \theta_1, \theta_2, \ldots \theta_N, 0)$ satisfies the requirement that $\theta_1, \ldots, \theta_K \in S$. In other words, the route begins from city 0 and needs to go through all cities defined by $S$ before visiting other cities.

For a route $\theta$, another route $\theta'$ is considered a *neighbor* route if the edit distance between $\theta$ and $\theta'$ is smaller than a certain threshold ($T$). For simplicity, only substitution is allowed in calculating the edit distance between $\theta$ and $\theta'$. In other words, a neighbor route $\theta'$ comes from changing the position of at most $T$ cities in $\theta$.

In the simulation experiment, the difficulty is adjusted by changing the value of K. Here we calculate the average time used to generate a block in the last 10 blocks. If it is greater than 60 seconds, we decrease K by one; if it is smaller than 15 seconds, we increase K by one.

The results of the simulation can be found on Figure 5. The best solution found in the first # of blocks generated so far, and the exact solution found using Held-Karp algorithm ("global minimum") are shown. Since the PoW only finds a best route up to the first K cities, we further optimized the route by iteratively enumerating the neighbor routes. The value of $T$ is set as 5, and a total of 3 miners compete to generate the block.

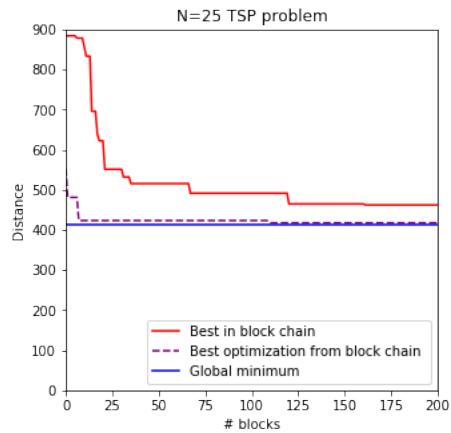

**Figure 5. The best route found in block chain simulation (red), an optimized route based on the best route (purple), and the exact solution (global minimum, blue), for the TSP problem in Figure 4.**

Figure 5 demonstrates that the best route in block chain (red line) quickly approaches the exact solution in the first 50 blocks generated, and a simple optimization based on the best route in block chain (purple line) quickly reaches the exact solution.

The time used to generate these blocks, and the value to adjust difficulty (K), is shown in Figure 6. After 20 blocks, the average time quickly reaches a desired value (between 15 - 60seconds), and the value of K remains 21 for the rest of the blocks, a demonstration of the success of our difficulty adjustment scheme.

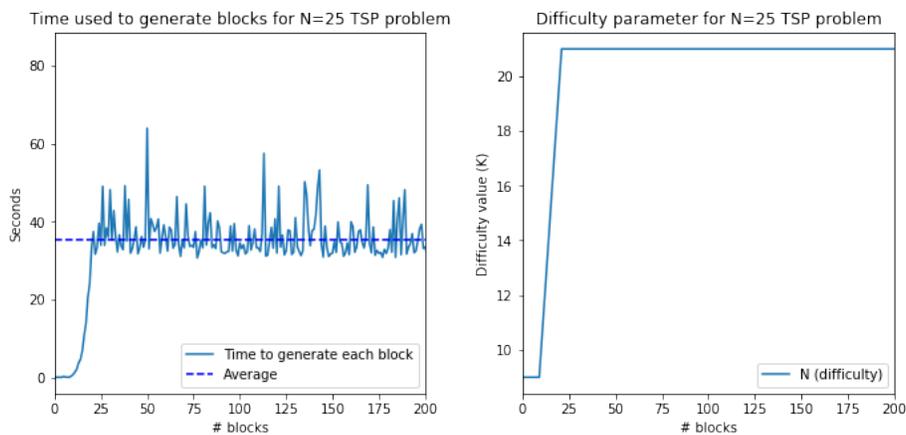

**Figure 6. The time (left) and difficulty value (K) used to generate each block.**

**Conclusion**
We revise the Proof-of-Work (PoW) algorithm commonly used in blockchain technologies to enable scientific computing. The revised PoW takes any form of high-dimensional, non-linear objective functions that are difficult to optimize but easy to validate. An index function is also introduced to enable incorporation of block information. By choosing the degree of freedom for optimization, the difficulties of PoW can be adjusted. We demonstrate the application of this PoW by solving a NP-hard Traveling Salesman Problem (TSP). Taken together, this PoW scheme helps solve a scientific question that generally requires a lot of computational resources.

There are several issues that need further investigation. First, the PoW validation checks whether a proposed solution is a BLM solution. This process may be time-consuming for high dimensions or larger BLM windows. To reduce the computational burden and speed up the process, all miners in the network may share the validation results. Once a counter example is found, it should spread over all miners in the network to reject the proposed solution. Second, a more complicated simulation is needed to demonstrate this framework is capable of handling frauds, and is able to solve a more meaningful problem (e.g., protein structure prediction).